\documentclass[12pt,preprint]{aastex}

\shorttitle{Cyclic behavior of solar inter-network magnetic field}
\shortauthors{Jin \& Wang}

\begin{document}

\title{Solar cycle variation of the inter-network magnetic field}

\author{Chunlan Jin \& Jingxiu Wang}
\affil{Key Laboratory of Solar Activity, National Astronomical Observatories,
 \ Chinese Academy of Sciences, Beijing 100012, China (cljin@nao.cas.cn)}

\begin{abstract}
Solar inter-network magnetic field is the weakest component of solar magnetism, but contributes most of the solar surface magnetic flux. The study on its origin has been constrained by the inadequate tempo-spatial resolution and sensitivity of polarization observations. With dramatic advances in spatial resolution and detective sensitivity, solar spectro-polarimetry provided by the Solar Optical Telescope aboard Hinode in an interval from solar minimum to maximum of cycle 24 opens an unprecedented opportunity to study the cyclic behavior of solar inter-network magnetic field. More than 1000 Hinode magnetograms observed from 2007 January to 2014 August are selected in the study. It has been found that there is a very slight correlation between sunspot number and magnetic field at the inter-network flux spectrum. From solar minimum to maximum of cycle 24, the flux density of solar inter-network field is invariant, which is 10$\pm1$ G. The observations suggest that the inter-network magnetic field does not arise from the flux diffusion or flux recycling of solar active regions, thereby indicating the existence of a locally small-scale dynamo. Combining the full-disk magnetograms observed by SOHO/MDI and SDO/HMI in the same period, we find that the area ratio of the inter-network region to the full-disk of the Sun apparently decreases from solar minimum to maximum but always exceeds 60\% even though in the phase of solar maximum.
\end{abstract}


\section{Introduction}

The complexity and irregularity of solar inter-network (intranetwork, inner-network, abbreviated as IN) magnetic field has been puzzling solar physicist. Its spatial scale ranges from a few arc-seconds down to the current telescope resolution limit, i.e., 0.1 arc-second. It is ubiquitous on the solar surface, just like the spreading `spiced salt'. As the smallest-scale magnetic structures on the Sun, the IN field rapidly appears and disappears, and provides a large fraction of the magnetic flux and energy to the solar photosphere at any given time (Stenflo 1982; Zirin 1987; Wang et al. 1995; S\'{a}nchez Almeida 2003; Trujillo Bueno et al. 2004). Solar community has paid more attention to the IN field due to its possible importance in understanding the structure of the quiet corona (Schrijver \& Title 2003), the chromospheric heating (Narain \& Ulmschneider 1996; Goodman 2004), the sources of the solar wind (Woo \& Habbal 1997; Tu et al. 2005). Furthermore, our understanding of IN field has been greatly improved by the observations with high spatial resolution and polarization sensitivity, e.g., the observation provided by Solar Optical Telescope (SOT; Tsuneta et al. 2008; Suematsu et al. 2008; Shimizu et al. 2008; Ichimoto et al. 2008) aboard Hinode (Kosugi et al. 2007). Our understanding will be continuously improved as more observations are analyzed, new instruments will come into service (e.g., ATST and Solar-C), and elaborated three dimensional magnetohydrodynamics numerical simulations of small-scale dynamo are getting progress.

To learn the physics of IN magnetic field is an important issue in solar physics. How and where does the IN magnetic field generate? What is the origin of the IN field? The answers to these questions are still controversial and debatable. At present, there are two main viewpoints about the source of IN field.
\begin{enumerate}
\item Small-scale dynamo. Petrovay \& Szakaly (1993) pointed that the observed IN field is only consistent with the model that it is generated and sustained by the vigorous convection very close to the solar surface, i.e., small-scale dynamo. Subsequently, the first magneto-convection numerical simulation of a fast dynamo indicates that the thermally driven turbulent convection can indeed be an extremely effective source of small-scale, highly intermittent magnetic field (Cattaneo 1999), which supports the idea that the solar granulation can act as the efficient dynamo. Furthermore, the relationship between the IN magnetic structures and the granular convection features is also observed (Lin \& Rimmele 1999; Khomenko et al. 2003; Lites et al. 2008; Jin et al. 2009a). Interestingly, the realistic simulation of local dynamo from Sch\"{u}ssler \& V\"{o}gler (2008) has obtained a close result found by Lites et al.(2008) based on Hinode Spectro-polarimetric (SP) data observation. Furthermore, the observations with high spatial resolution and polarization sensitivity have provided the necessary supportive condition for the operation of the small-scale dynamo (Lites 2011).

\item Flux of decaying active region. Some fractions of quiet magnetic fields are likely to result from flux diffusion of decaying active region as evidenced by the similar 'butterfly diagram' obtained by Jin et al. (2011) and Jin \& Wang (2012), but most of the magnetic signals in the butterfly diagram are the stronger network structures but not the IN magnetic structures. Another possible explanation of IN magnetic structures is that they result from the interaction of convection with the magnetic flux of decaying active regions. The magnetic flux from the decaying active region is likely to be dragged down by the convection and then emerge again on the solar surface, i.e., flux recycled (Ploner et al. 2001; de Wijn et al. 2005).
\end{enumerate}

Considering the fact of cyclic dependence of decaying flux from active region that diffuses into IN regions, one must expect that there should be a correlation between the unsigned magnetic flux density of IN regions and sunspots if the IN magnetic field arises from the magnetic flux of decaying active regions. The advances in tempo-spatial resolution and polarization sensitivity of SP observation provided by SOT on board Hinode provide the possibility to identify the very weak magnetic signals, thereby to study the cyclic behavior of solar IN magnetic field based on the long-term observation. Buehler et al. (2013) examined the cyclic variations of the linear and circular polarization signals detectable at the disk center of the quiet Sun based on the Hinode/SOT observations, they found that the fraction of pixels with significant polarization signals remained constant from 2006 to 2012. Recently, Lites et al. (2014) explored the latitude-time behaviour of the weak magnetic signals observed by the Hinode SOT. They found no evidence for obvious changes as a function of time and solar latitude in the unsigned line-of-sight field and the transverse field.

The distinction between IN region and network region is subjective in some sense and dependent on the spatial resolution and polarization sensitivity of the telescope. However, the more the reasonable constraint conditions are, the more accurate the extraction of IN region should be in a way. Comparing the IN region extracted by mainly considering the pixels' polarization signal of Buehler et al. (2013) and Lites et al.(2014), we distinguish the IN magnetic field from network field by considering both the polarization signal and the area of magnetic structures in this paper. The aim is to explore the cyclic behavior of IN magnetic field based on the identification of the IN magnetic structures, and analyze the correlation between sunspot number and IN magnetic flux. To make a relatively safer identification of IN magnetic structures before studying their cyclic variation would guarantee a safer conclusion. In this study we report that from solar minimum to the current maximum of cycle 24, the magnetic flux density of solar IN field does not show variation beyond the scope of measurement noise, though the area occupied by IN field apparently decreases during the same interval. Furthermore, the correlation between sunspot number and magnetic field at IN flux spectrum is very low. These results indicate that the solar IN field comes from a small-scale dynamo which does not correlate with sunspot cycle. The next section is devoted to the description of the observations and data analysis. The results are presented in Section 3. In section 4, the conclusions are drawn and discussions are presented.

\section{Observations and data analysis}

The SOT/SP measurements provide high spatial resolution and polarization sensitivity observations since 2006 November. We totally selected 1024 magnetograms, which cover the period from 2007 January to 2014 August, including solar minimum and the current maximum of solar cycle 24. The detailed monthly distribution of the magnetogram number is presented by a histogram in Figure 1. From the figure, it can be found that there are more than 600 magnetograms for quiet Sun, which are shown by red box and distributed principally in the years of 2007-2009; there are almost 200 magnetograms for active regions, which mainly spreads the year of 2014. In addition, about 200 magnetograms of enhanced network regions or plage regions are also selected throughout the entire period. We avoid the regions close to the solar limb and polar region by selecting magnetograms with distance less than 60 degrees from solar disk center.

The integration time for these selected magnetograms is at least 3.2 s, and the number of the fast map (i.e., the map with 3.2 s integration time) occupies 78\% of the total. In this study, we mainly focus on the weak polarization signals. In order to enhance the sensitivity of weak polarization signals in the presence of measurement noise, the measurement of line-of-sight magnetic field is based on the wavelength-integrated method, which has been described in detail by Lites et al. (2008) and Jin et al. (2009b). The stronger polarization signals in the quiet Sun are inverted to determine the calibration constant relating the line-of-sight field and the integrated circular polarization. The method avoids these problems of non-convergence and non-uniqueness that arise in the inversions of noisy profiles, and pushes the analysis of very weak observed signals. However, there is an intrinsic weakness of the method. It does not give a correct estimate for the strong field with intrinsic value larger than kilo-gauss because of the saturated profiles (Lites et al. 2008). In this study, such strong fields do not make significant influence on the main results of the analysis. Because the observing modes are non-uniform for these selected magnetograms, their magnetic noise levels are estimated, respectively. We adopt two approaches to estimate the magnetic noise: (1) we select a quiet region with very weak polarization signal, and compute the standard deviation as the magnetic noise; (2) we analyze the histogram of flux densities within the magnetograms themselves, and then describe the core of the distribution function by a Gaussian function, and adopt the half width of the Gaussian function as the noise. There is little difference of magnetic noises obtained by the two approaches. For different observing modes, the magnetic noises fall in the range of 2.5 G to 3.5 G. Here, we adopt the uniform noise level of 4 G.

Just as described above, not all of these selected magnetograms are located in the solar disk center, but in the range of 0 to 60 degrees far away from solar disk center. More importantly, the earlier studies have discovered the limb weakening of the line-of-sight magnetic field while the transverse field almost keeps constant with far away from solar disk center (e.g., Jin \& Wang 2011). The origin of the limb weakening of line-of-sight magnetic field comes from not only the geometry of an expanding flux tube but also the varying formation height affected the radiative transfer in the stratified atmosphere. At the present time there is a uniform correction of the limb weakening. According to simulated result of Jin et al. (2013) by using the thin flux tube model (Fontenla et al. 2006, 2007), the variation of the observed line-of-sight field conforms to the cosine function with far away from solar disk center, so the observed line-of-sight component is corrected as $B_{cal}=B_{obs}(\alpha)/cos(\alpha)$. The angle $\alpha$ is defined by $sin(\alpha)=\sqrt{x^2+y^2}/R$, where $x$ and $y$ are the pixel position referring to the solar disk center, and the $R$ is the solar disk radius. In this paper, we adopt the above method to correct the limb weakening of the line-of-sight field. In addition, the pixel area of all magnetograms is also corrected according to the cosine function.

In order to compute the area occupancy of IN regions to the full-disk magnetogram in this period, the full-disk line-of-sight magnetograms obtained from the Helioseismic and Magnetic Imager (HMI; Scherrer et al. 2012) on board Solar Dynamics Observatory (SDO; Pesnell et al. 2012) and the Michelson Doppler Imager (MDI) on board the Solar and Heliospheric Observatory (SOHO; Scherrer et al. 1995) are used. The 5-min average magnetograms of MDI are selected in this study, which include the period from 2007 January to 2010 December. The full-disk magnetograms from 2011 January to 2014 August are taken from HMI observations. In order to improve the sensitivity of HMI magnetogram, seven continuous HMI magnetograms per day are selected and then averaged in time. Therefore, all full-disk magnetograms from MDI and HMI are the 5-min average observation.

For these full-disk magnetograms observed by MDI and HMI, we firstly exclude the active regions by creating a logical mask, and this method has been described in detail by Jin et al. (2011) and Jin \& Wang (2012, 2014). Therefore, the area occupancy of quiet region to the full-disk Sun is computed. For the SP magnetograms, we firstly exclude the active regions by the same method. An example is displayed in the middle panel of Figure 2. Comparing the top panel with the middle panel in Figure 2, it can be found that the active region and surrounding magnetic environment have been excluded by our method. In the magnetogram, the area occupancy of quiet Sun reaches 90.8\% of the entire field-of-view area. For these SP magnetograms after excluding active regions, we identify all magnetic structures above the noise level of 4 G and the area of 0.2 Mm$^{2}$ (i.e., the area of the magnetic structure is larger than 4 pixels if the spatial scale of pixel is 0.32 arcseconds, and larger than 16 pixels if the spatial scale of pixel is 0.16 arcseconds). The method of identifying magnetic structure has been described in detail by Jin et al. (2011). Then we compute the magnetic flux of each magnetic structure. In order to distinguish the network region and IN region, a logical mask is created by excluding the magnetic structures with magnetic flux larger than 10$^{18}$ Mx. According to the flux distribution of IN and network magnetic structures (Wang et al. 1995; Zhou et al. 2013), we cannot exclude the possibility that some weaker network magnetic structures are still remained in the studied IN regions. However, at the present time, we cannot find a better method to distinguish IN regions from network regions based on the magnetic distribution. An example of identifying IN region is shown in the bottom panel of Figure 2. It can be found that the stronger magnetic structures have been excluded by distinguishing their magnetic flux, and the area occupancy of IN region to the entire magnetogram reaches 60.4\%.

\section{Results}

The aim here is to quantify the variation of the solar IN magnetic field with very weak polarization signals. If the flux in the IN region results primarily from the diffusion or the recycling of decaying active region, one would expect to see a corresponding variation of IN field from solar minimum to maximum.

We compute the magnetic flux density of IN regions in each selected SP magnetogram by considering the magnetic noise level of 4 G. Then the monthly average magnetic flux density of IN region is computed. The black diamond symbol in the middle panel of Figure 3 displays the monthly average variation of flux density in the IN magnetic field during the interval of 7.67 years, and the corresponding variation of monthly sunspot number is shown in the bottom panel. There is no cyclic variation seen in the IN flux distribution, i.e., the magnetic flux density of IN magnetic field is 10 $\pm$1 G during the ascending phase of solar cycle 24. The average magnetic flux density in solar inter-network region has been reported earlier to be 11 G by Lites et al. (2008). They analyzed a SP magnetogram taken on 2007 March 10, and derived the average line-of-sight field of about 11 G from wavelength-integrated measurements. Our result verifies the universality of inter-network magnetic flux density of 10 $\pm$1 G, especially the universality in entire ascending phase of solar cycle 24.

It has been clarified that magnetic features with different size and flux on the quiet Sun show different cyclic behaviors in term of their correlation to sunspot numbers in solar cycle (Hagenar et al. 2003; Jin et al. 2011). Without considering the flux spectrum of Sun's small-scale magnetic field, it would be no doubt that small-scale magnetic structures as a whole follow the sunspot cycle in phase. The reason is that the majority of small-scale field which dominant the total flux are debris from sunspot field. We have studied the cyclic variation of magnetic flux spectrum in quiet Sun, and found that from the detective limit of MDI (i.e., 10$^{18}$ Mx) to the large end of flux spectrum, the variations of magnetic structure number show no correlation, anti-correlation and correlation with sunspot number. The correlated network magnetic structures are likely to be the debris of decayed sunspots. The anti-correlated magnetic structures are the main discovery in the work of Jin et al. (2011), and the physical explanation of their source needs the help with the sophisticated MHD numerical simulation and more observations, but some possibilities are considered to explain the anti-correlated magnetic structures in term of the interplay of local and global dynamos (Jin et al. 2011). It is worthwhile mentioning, the anti-correlation of the small-scale solar activity, e.g., X-ray bright points, with the sunspot number changes in a solar cycle has been known for quite long time (e.g., Golub et al. 1979; Davis 1983; Muller \& Roudier 1984, 1994; Harvey 1985). The discovery of an anti-correlated component of solar magnetogram may shed new light for a satisfactory understanding on the small-scale solar activity. While the no-correlated magnetic structures lie in the detective limit of MDI and their number is very few, we cannot exclude the possibility that the no-correlated magnetic structures come from the measurement noise. With the improvement of the spatial resolution and polarization sensitivity, at present the magnetic flux of the observable smallest magnetic structures reaches 10$^{16}$ Mx, and it is necessary and possible to examine the cyclic variation of these magnetic structures lying in the detective limit of MDI. In this paper, based on the observations from Hinode/SP, we analyse the cyclic variation of the magnetic flux spectrum from 10$^{18}$ Mx down to 10$^{16}$ Mx, which mainly consists of IN magnetic structures. For these magnetic structures, we divide them into 19 bins according to their magnetic flux magnitude, and compute the monthly average magnetic flux density in each bin based on more than 7 years observations. 

In such a way, the cyclic variation of magnetic flux density in each bin is obtained, and the correlation coefficients between the flux density and sunspot number are computed, which is shown in Figure 4. \textbf{Here the correlation coefficient refers to the Spearman¡¯s rank correlation. The correlation coefficients between the sunspot number and the magnetic flux of magnetic structures with flux larger than 10$^{18}$ Mx are little different from those obtained in Jin et al. (2011). Jin et al. (2011) computed the linear Pearson correlation coefficient. The procedure to calculate the Spearman's correlation provides both the correlation coefficient and confidence level. From the statistic point of view, both the correlation coefficients indicate high correlation. In certain scope of uncertainty they are consistent.} The color bar shown in Figure 4 represents the confidence level of the correlation, and a large value of the confidence level indicates a significant correlation. From the smallest end to the 10$^{18}$ Mx of the flux spectrum, there appears obviously no correlation between the IN magnetic structure and the sunspot. In summary, in this study the lack of substantial correlation between IN field and sunspot number in the process of cycle 24 suggests that the IN magnetic field is independent of the sunspot magnetic field. In other words, the decaying flux of active region is not likely to be responsible for the production of IN magnetic field. This indicates that the operation of a small-scale dynamo is probably the source of IN magnetic flux in the solar cycle.

Using these SP magnetograms, the area occupation of IN regions to the entire quiet field-of-view is computed by distinguishing IN region from network region. We assume that the area occupation of IN region in local Sun is consistent with that in full-disk solar surface. Then combined the area ratio of quiet region to the entire solar surface obtained from HMI and MDI full-disk magnetograms, the area ratio of the IN regions to the full-disk Sun is obtained, which is shown in the top panel of Figure 3. It can be found that the IN area gradually decreases with increasing sunspot number, and the area ratio of IN region exceeds 60\% even though in the period of the solar maximum.

\section{Conclusion and Discussions}

Examination of the weak polarization signals (i.e., for the IN regions) in 1024 solar magnetograms obtained by the SOT/SP aboard Hinode from 2007 January to 2014 August excludes the possibility that IN magnetic field mainly arises from the magnetic flux of decaying active regions. The IN magnetic flux density is a constant within the scope of noise level from solar minimum to the current maximum of cycle 24. Furthermore, the magnetic field at IN magnetic flux spectrum shows very low correlation with sunspot number, which confirmed the earlier results (Buehler et al. 2013; Lites et al. 2014). The observations suggest that the IN magnetic field is produced by a small-scale dynamo independent of sunspot cycle.

We recognized that some improper data analysis would affect the result. At present, there is no a uniform correction of the limb weakening of the line-of-sight field, and the reliability of the cosine function correction still needs to be examined. Here we use the an empirical correction of line-of-sight magnetic field to reanalyze the cyclic behavior of IN magnetic flux density. Considering the larger flux density in active regions than that in quiet regions and the cyclic dependence of quiet magnetic fields (Jin et al. 2011), we first select 277 quiet magnetograms covering the period from 2007 September to 2009 August during which there were less sunspots and no obvious cyclic variation. Secondly the magnetic flux density in these quiet regions is computed. The correlation between flux density and heliocentric distance is given, illustrated in the top panel of Figure 5. Finally the correction factor of flux density with heliocentric distance is given by setting the sort-group bin and fitted by a polynomial function, which is shown in the bottom panel of Figure 5. The correction factor is used to all magnetograms selected in this study to correct the line-of-sight magnetic field, and the magnetic flux density in IN region is computed. The corresponding variation of IN magnetic field is shown by the red triangle symbol in Figure 3. It can be found that there is no obvious difference from previous result except for the relatively smaller values of the IN flux density.

Besides the influence from projection effect correction, there is another possible procedure which would possibly affect the cyclic variation of IN magnetic field: the randomness of local SP observation and our selection of the magnetograms. Sometimes one region was close to strong magnetic field, e.g., an active region, but another region was within a large-scale quiet region. Therefore they might suffer from different influences from either strong or weak neighbourhood flux distribution by some types of flux diffusion. To learn whether or not this effect would play a role in the observed IN flux density for selecting regions, we compare the different types of regions but observed in the same cycle phase, and find that their magnetic flux densities in the IN region are almost the same. Just as shown in the comparing example of Figure 6, the measurements of quiet and active regions are observed in the same month, and magnetic flux density is 15 G for entire quiet map, and 51 G for the magnetogram including the active region. Considering the noise level of about 4 G, the IN magnetic flux densities in the two measurements are both 11 G. This suggest that the magnetic flux density of IN region is independent of the neighbouring flux density distribution, and the information of the local measurement from these selected SP magnetograms basically represents the content of IN region in full-disk magnetogram.

\acknowledgments

The authors are grateful to the team members who have made great contribution to the SDO mission and Hinode mission. SDO is a mission of NASA's living with a Star Program. Hinode is a Japanese mission developed and launched by ISAS/JAXA, with NAOJ as domestic partner and NASA and STFC (UK) as international partners. It is operated by these agencies in co-operation with ESA and NSC (Norway). The work is supported by the National Basic Research Program of China (2011CB811403) and the National Natural Science Foundation of China (11003024, 11373004, 11322329, 11221063, KJCX2-EW-T07, and 11025315).

\begin{figure}
\includegraphics[scale=0.8]{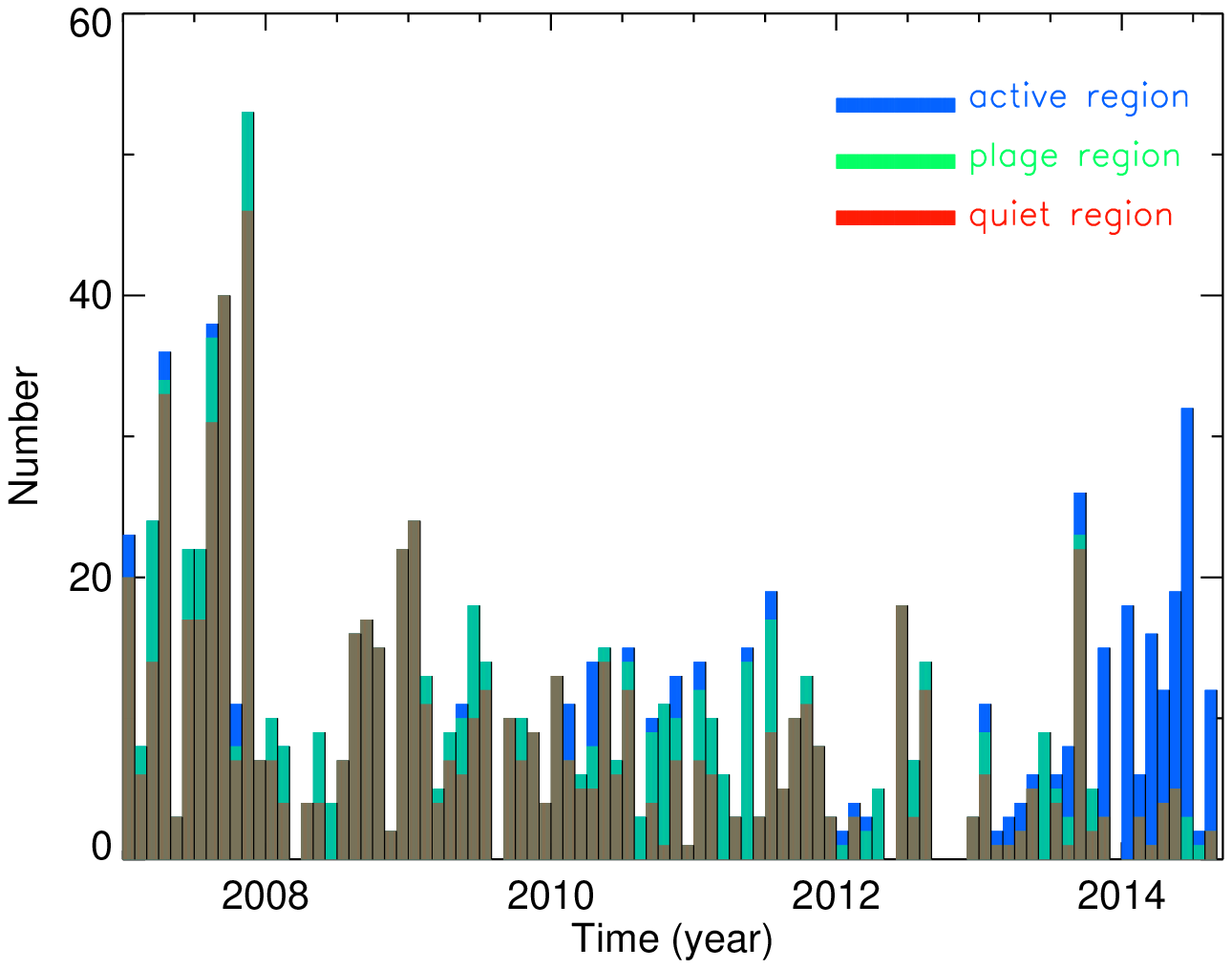}
\caption{The monthly distribution histogram of the selected magnetogram number. The numbers of active region, plage region and quiet region are shown by blue box, green box and red box, respectively.}
\end{figure}

\begin{figure}
\includegraphics[scale=0.8]{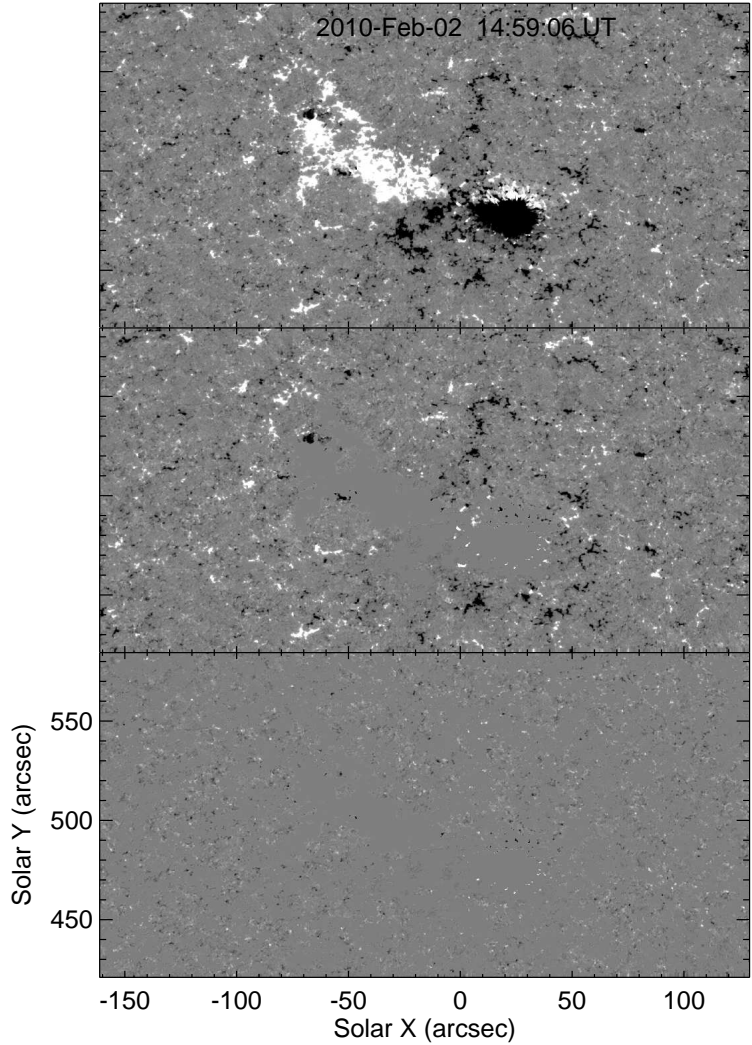}
\caption{The extraction of IN region. Top: the line-of-sight magnetic field distribution for Hinode/SP data recorded 2010 February 02 beginning at 14:59 UT. Middle: the quiet magnetogram after excluding the active region. Bottom: the IN region excluding the magnetic structures with magnetic flux larger than 10$^{18}$ Mx based on the quiet magnetogram.}
\end{figure}

\begin{figure}
\includegraphics[scale=0.8]{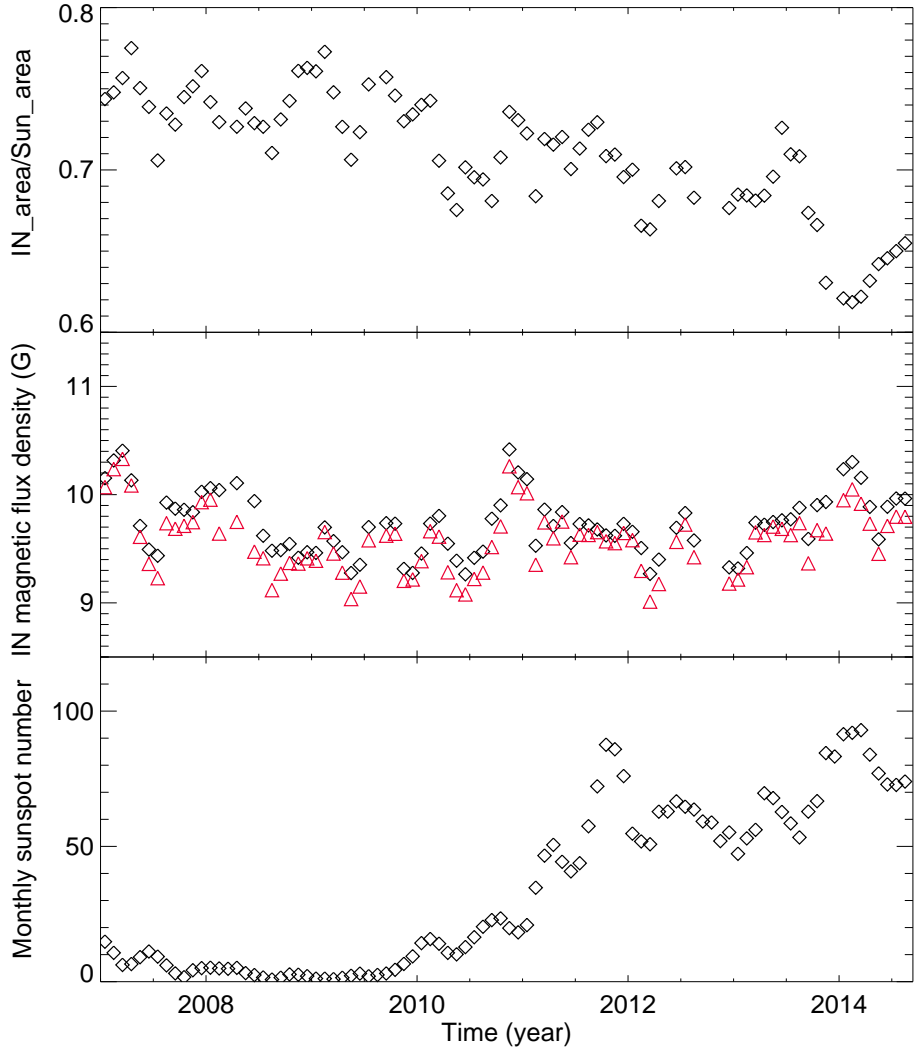}
\caption{The cyclic variation of IN magnetic field. Top: the area occupation of IN region to the full-disk Sun. Middle: the distribution of IN magnetic flux density, where the black diamond symbol means the result computed by the model correction of line-of-sight magnetic field, and the red triangle symbol represents that obtained by empirical correction of line-of-sight magnetic field. Bottom: the corresponding variation of monthly sunspot number in this period.}
\end{figure}

\begin{figure}
\includegraphics[scale=0.8]{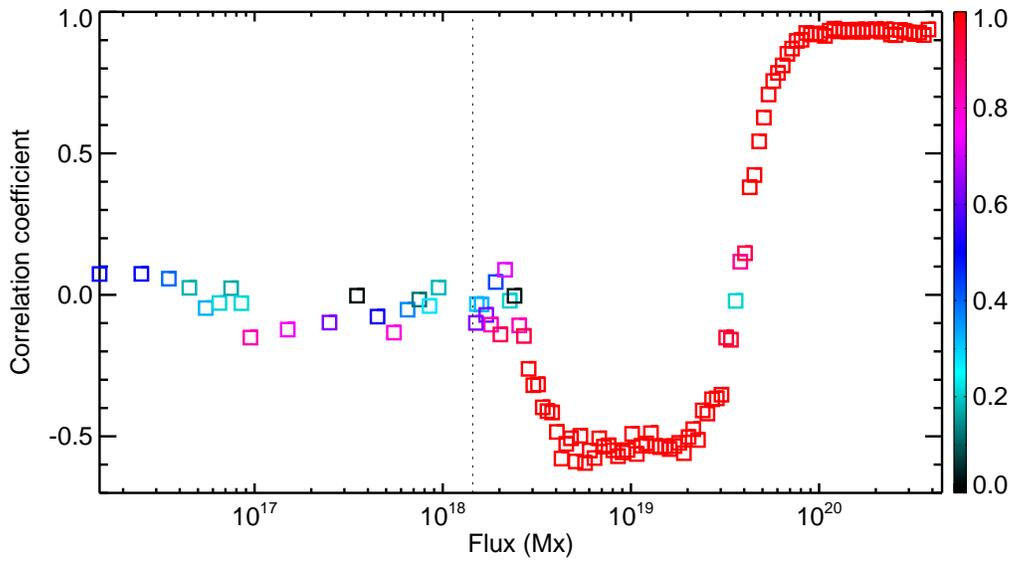}
\caption{The correlation coefficients between sunspot cycle and cyclic variation of flux density of magnetic structures with different fluxes. The color bar represents the correlation confidence level. The vertical dotted line distinguishes the flux of magnetic structure observed by MDI from Hinode/SP.}
\end{figure}

\begin{figure}
\includegraphics[scale=0.8]{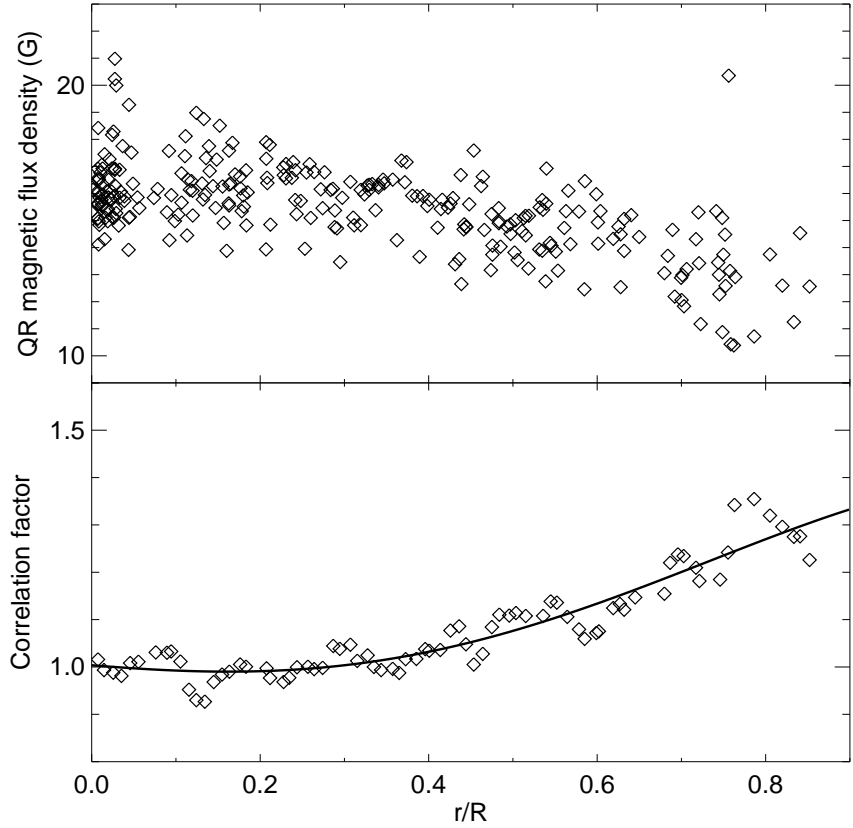}
\caption{The empirical correction of line-of-sight magnetic field from solar disk center to the limb. Top: the magnetic flux density distribution in quiet region with far away from solar disk center, where the quiet magnetograms are observed from 2007 September to 2009 August considering the cyclic dependence of quiet magnetic field and larger magnetic flux density in active region than in quiet region. Bottom: the correction factor of line-of-sight magnetic field obtained by sort-group bins according to the distance from solar disk center, where the black solid line means a polynomial fitting.}
\end{figure}

\begin{figure}
\includegraphics[scale=0.8]{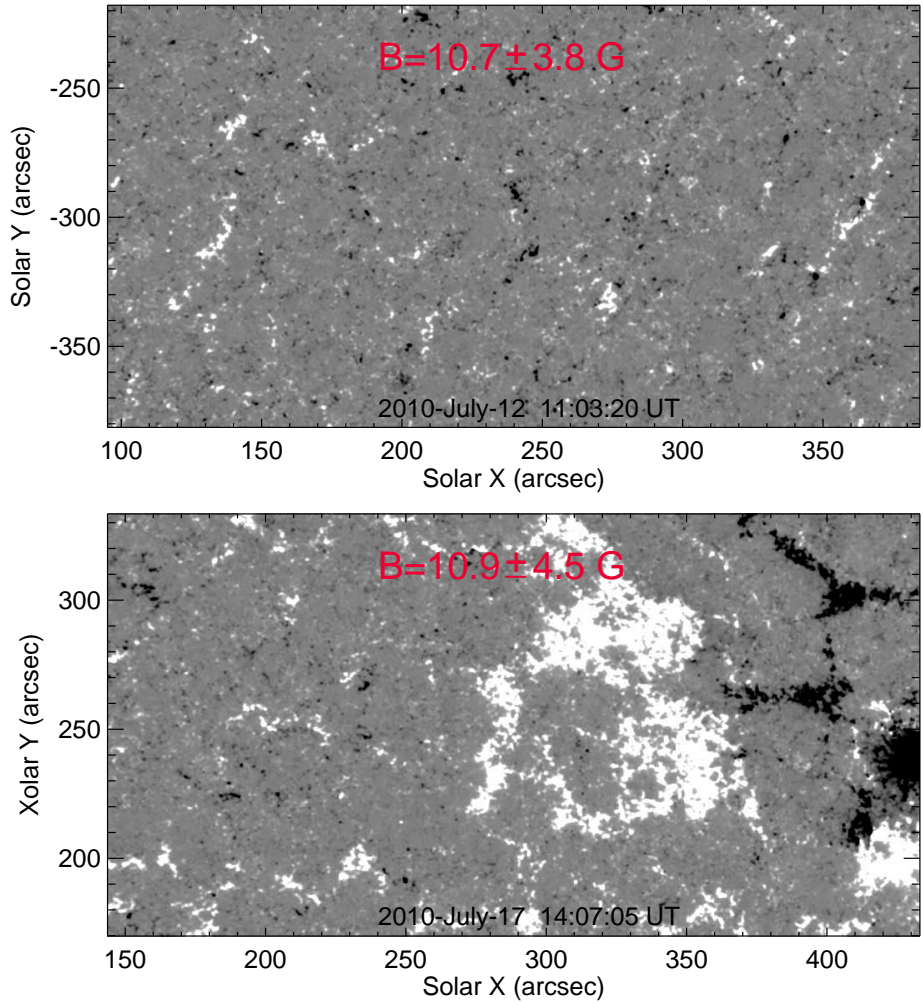}
\caption{A comparing example of the IN regions locating in different types of flux distributions but in the same phase. The magnetic flux density reaches 15 G in the quiet magnetogram, and 51 G in the active magnetogram. However, the IN flux densities are the same in the two magnetograms.}
\end{figure}

\end{document}